\documentclass[12pt]{article}
\usepackage{amsfonts}
\usepackage{amssymb}
\usepackage{graphicx}
\usepackage{amsmath,graphicx,amsfonts}
\usepackage{epsfig}
\usepackage{subfigure}
\usepackage{dcolumn}

\def\siz{\small}

\def\be{\begin{equation}}
\def\ee{\end{equation}}

\begin{document}
\title{Self-gravitating darkon fluid with anisotropic scaling}
\author{P.C. Stichel$^{1)}$  and W.J.
Zakrzewski$^{2)}$
\\
\siz
$^{1)}$An der Krebskuhle 21, D-33619 Bielefeld, Germany \\ \siz
e-mail:peter@physik.uni-bielefeld.de
\\ \\ \siz
$^{2)}$Department of Mathematical Sciences, University of Durham, \\
\siz Durham DH1 3LE, UK \\ \siz
 e-mail: W.J.Zakrzewski@durham.ac.uk
 }

\date{}
\maketitle

\begin{abstract}
The fluid model for the dark sector of the universe (darkon fluid), introduced previously in \cite{PRD}, is
 reformulated as a modified model involving only variables from physical phase space. The Lagrangian 
of the model does not possess a free particle limit and hence the particles it describes, darkons,
exist only as a self-gravitating fluid. This darkon fluid presents a dynamical realisation of the 
zero-mass Galilean algebra extended by anisotropic dilational symmetry
with dynamical exponent $z=\frac{5}{3}$. The model possesses cosmologically relevant solutions 
which are identical to those of \cite{PRD}. We derive also the equations for the cosmological
perturbations at early times and determine their solutions. In addition, we discuss also some implications of adding higher spatial-derivative  terms.

\end{abstract}

\section{Introduction}

In the previous paper \cite{PRD} we have presented a model of dark energy based on new particles
which, in the following, we call ``darkons". These particles, which are nonrelativistic and massless,
 possess a modified relation between energy and momentum (or velocity) and so in their description 
we were forced to use an enlarged phase space (see \cite{us}). The gravitational coupling of the
darkons was introduced 
 in a minimal way which satisfied the general form of the 
Einstein equivalence principle. This minimal coupling  lead to a dynamically generated active gravitational mass density of either sign as a source of the gravitational field. Such a property opened the possibility of using this model to explain the observed accelerated expansion of the universe (see \cite{three} for a very recent review).
Our paper \cite{PRD} then used a fluid mechanical generalisation of this particle concept to construct a new parameter free model for the dark sector of the universe. And, when we
performed a comparison of the predictions of this model with the cosmological observations, we were
pleased to see no disagreement thus suggesting that we should look at it more seriously.
 
The enlarged phase space, which 
was used in the construction of this model, seems somewhat unphysical. Hence we have decided to see whether we can reformulate the model using only the conventional 
phase space. This paper presents such a reformulation.

The second important property of our model is the anisotropic scale symmetry defined by the following transformation properties of space and time coordinates:
\begin{equation}
\label{ex1}
\vec x^{\star}\,=\,\lambda \vec x,\qquad t^{\star}\,=\,\lambda^z t,
\end{equation}
where $z$ is a dynamical exponent. Any relativistic scale invariant theory clearly shows isotropic scaling ({\it i.e.} $z=1$) as well as possesses the corresponding nonrelativistic limit \cite{LSZ}.
Galilei covariant theories with nonvanishing mass can exhibit the anisotropic scaling with $z=2$ 
\cite{Hagen}, the so-called Schr\"odinger symmetry. Galilei covariant theories with $z\ne 2$ have necessarily the vanishing mass \cite{us} and so this is the case also for our model, which as we
will demonstrate, corresponds to $z=\frac{5}{3}$. Such a $z$-value for a gravitational theory
is exceptional. Recent renormalizable gravitational theories, which violate Poincar\'e symmetry in the ultraviolet limit and which exhibit anisotropic scale symmetry there, usually correspond to $z=3$ \cite{Horava}.
Cosmologies with other $z$-values have also recently been discussed (see \cite{Yu}).

But there is a crucial difference between Horava's gravity \cite{Horava} and our work in the way they get $z\ne1$. In \cite{Horava} higher-order spatial derivative terms corresponding to $z=3$ are introduced from the very beginning - in order to enforce renormalizability of the theory. In our case, however, the anisotropic scaling with $z=\frac{5}{3}$ is an emerging symmetry resulting from the minimal coupling of massless nonrelativistic particles to gravity.

For any model, to be a candidate for the description of the universe, one should determine, in addition
to its cosmological solution, also the perturbations around this solution. This is what we do for our model in this paper.
Basing our discussion on the symmetry properties of our model we derive perturbations around its
cosmological solution at early times and present a preliminary discussion of their form.

The paper is organised as follows. In Sect. 2 we review the previous formulation of our model and describe the modifications in its description so that the unphysical degrees of freedom used in its 
original formulation are absent. In Sect. 3 we show that the darkon fluid is an interacting system
and, as such, possesses zero mass Galilean symmetry which will be extended to anisotropic scaling
in Sect. 4. In Sect. 5 we derive the equations for cosmological perturbations at early times 
and describe their solutions. Finally, in Sect. 6, we add higher spatial-derivative terms to the interaction 
and discuss their implications. We conclude with some remarks and an outlook for further research in this area (Sec. 7).

\section{The reformulation of the model}

\subsection{Previous formulation}
The two-component fluid model in \cite{PRD} is defined in terms of a Lagrangian given by

\begin{equation}
\label{eten}
L\,=\,L_M\,+\,L_D\,+\,L_{\phi},
\end{equation}
where $M$, resp. $D$, stands for the baryonic resp. dark sector.

In \cite{PRD} we have started with another interpretation: $M$ for the matter (baryonic and dark) and $D$ for the dark energy sector. But as shown in \cite{PRD} the $D$ sector can be used for a unified
description of the whole dark sector. Hence, in the present paper, we will follow this attitude 
from the very beginning.

 For the separate parts of $L$ in (\ref{eten}) we have
\begin{equation}
\label{eeleven}
L_M\,=\,mn_0^M\int \,d^3\xi\,\left(y_i^M(\dot x_i^M\,-\,\frac{1}{2}y_i^M)\,-\,\phi(\vec{x}^M,t)\right),
\end{equation}
where $m$ is a mass parameter giving (\ref{eeleven}) the correct dimension,
\begin{equation}
\label{etwelve}
L_D\,=\,n_0^D\int\,d^3\xi\,\left(p_i(\dot x_i^D-y_i^D)\,+\,q_i^D\dot y_i^D\,+\,q_i^D\,\partial_i\,\phi(\vec{x}^D,t)\right)
\end{equation}
and
\begin{equation}
\label{ethirteen}
L_{\phi}\,=\,-\frac{1}{8\pi G}\,\int\,d^3x\,\left(\vec{\nabla}\phi(\vec{x},t)\right)^2,
\end{equation}
where $n_0$ denotes the constant particle distribution in $\vec \xi$ space ($\vec \xi$ is a continuous particle label) in the respective sector and $G$ is Newton's gravitational constant.
In these expressions all phase space variables are functions of $\vec{\xi}$ and $t$, {\it ie} $\vec{x}^M=\vec{x}^M(\vec{\xi},t)$ etc. Note that the variable $\vec q^D(\vec \xi,t)$ acts as a Lagrange 
multiplier field \footnote{For the role of Lagrange multiplier constraints in gravity theory see \cite{newtwo}.}.
For more details see \cite{PRD}.

From (\ref{etwelve}) we note that the phase space in the $D$ sector is now 12-dimensional.

The equations of motion (EOM), corresponding to $L$, are then given by
\begin{itemize}
\item $M$ sector
$$  \dot x_i^M\,=\,y_i^M$$
\begin{equation}
\label{efourteen}
\dot y_i^M\,=\,-\partial_i\,\phi(\vec{x}^M,t)
\end{equation}

\item $D$ sector
$$  \dot x_i^D\,=\,y_i^D$$
\begin{equation}
\label{efifteen}
 \dot q_i^D\,=\,-p^D_i
\end{equation}
$$ \dot y_i^D\,=\,-\partial_i\,\phi(\vec{x}^D,t)$$
$$  \dot p_i^D\,=\,q_k \partial_k\partial_i\,\phi(\vec{x}^D,t)$$
\item $\phi$ sector
\end{itemize}
\begin{equation}
\label{esixteen}
 \triangle \phi(\vec{x},t)\,=\,4\pi G \int \,d^3\xi\left(
mn_0^M\delta(\vec{x}-\vec{x}^M(\vec{\xi},t))\,+\,n_0^Dq_i(\vec{\xi},t)\,\partial_i\,\delta(\vec{x}-
\vec{x}^D(\vec{\xi},t))\right).
\end{equation}
The last term in (\ref{esixteen}) represents a dynamically generated active gravitational 
mass density of either sign leading either to an attractive or a repulsive gravitational force. For this
reason the $D-$ sector may serve as a model for the dark sector of the universe (for more details see 
\cite{PRD}).

All this discussion has been given in the Lagrange formulation. However, for the fluids 
it is more convenient to express all quantities in the Eulerian formulation.

In such a case the dynamics of a fluid is usually described by the fluid particle density field $n(\vec x, t)$,
the velocity field $\vec u(\vec x, t)$ and the components of the force fields.
These fields are given by

\be
n(\vec x,t)\,=\, n_0\int\, d^3\xi\,\delta^3(\vec x-\vec x(\vec \xi,t))
\label{ex}
\ee
and
\be
u_i(\vec x,t)\,=\,y_i(\vec \xi(\vec x,t),t),
\label{ex1}
\ee 
where $y_i(\vec \xi,t)$ is an independent degree of freedom and the functions $\vec \xi(\vec x,t)$ 
are the inverse of $\vec x(\vec \xi,t)$. 
In our case \cite{PRD}
we have, in addition to the $n$ and $\vec u$ fields, also 
the momentum $p_i(\vec x,t)$ and pseudo-momentum $q_i(\vec x,t)$ fields in the $D$ sector
which are given by $p_i(\vec \xi(\vec x,t),t)$, resp. $q_i(\vec \xi(\vec x,t),t)$.

The corresponding EOM are then given by (for more details see \cite{PRD})
\begin{equation}
\partial_t\,n^A(\vec x,t)\,+\, \partial_k(n^A u_k^A)(\vec x,t)\,=\,0, \label{a}
\ee
where $A\in (M,D)$, 
{\it ie} the continuity equations for the particle number densities $n^M$ and $n^D$.

The Poisson equation for the gravitational field is given by
\be 
 \triangle \phi(\vec{x},t)\,=\,4\pi G \left(\rho^M\,+\,\partial_i(n^Dq_i)\right),
\label{b}
\ee
where the mass density $\rho^M$ is defined by $\rho^M:=m n^M$.

Note that the last term in (\ref{b}) represents the dynamically generated active gravitational mass density of the 
dark fluid.

As shown in \cite{PRD} we have, in addition, the following equations:
\be
\label{c}
D_t^M\,u_i^M\,=\,-\partial_i\phi
\ee
from the second equation in (\ref{efourteen}) and from the third equation in
(\ref{efifteen})
$$ D_t^D\,u_i^D\,=\,-\partial_i\phi,$$
where we have defined 
\begin{equation}
\label{enew}
D_t^A=\partial_t + u_i^A\partial_i.
\end{equation}

Note that if $u_i^M$ and $u_i^D$ obey the same initial conditions  
(\ref{c}) shows that $u_i^D=u_i^M=u_i$ {\it ie} (\ref{c}) becomes one universal
equation valid for all fluid components.
\be
\label{d}
D_t\,u_i\,=\,-\partial_i\phi.
\ee

Finally, the second and fourth equations in (\ref{efifteen})  give
\be
D_t\,q_i\,=\,-p_i,\qquad D_t\,p_i\,=\,q_k\partial_i\partial_k\,\phi.
\label{e}
\ee
Looking at (\ref{d},\ref{e}) we note that, in contrast to standard fluid mechanics,
the two vector fields $\vec p(\vec x,t)$ and $\vec u(\vec x,t)$ are not parallel to each other.

\subsection{Modifications of the dark sector}
All this was introduced and discussed in \cite{PRD}. However, let us observe that the gravitational field $\phi(\vec x,t)$ only appears in  $L_D$ and $L_\phi$  in the expression for the gravitational force
$g_i=-\partial_i \phi$. This suggests to introduce $g_i$ as the field variable in the Lagrangian restricted to the dark sector and its self-interaction. Then,
instead of (\ref{esixteen}) we would have\footnote{As only the $D$ sector is considered we omit the superscript $D$ here and in the following.}
\begin{equation}
\label{N1}
g_i(\vec x,t)\,=\,-4\pi G n_o\,\int\,d^3\xi\,\delta^3(\vec x-\vec x(\vec \xi,t))\,q_i(\vec \xi,t),
\end{equation}
which can be rewritten as
 
\begin{equation}
\label{N2}
g_i(\vec x,t)\,=\,-4\pi G n(\vec x,t)\, q_i(\vec \xi(\vec x,t),t),
\end{equation}
and so
\begin{equation}
\label{N3}
q_i(\vec \xi,t)\,=\,-\frac{1}{4\pi G} \left( \frac{g_i}{n}\right) \left(\vec x(\vec \xi,t),t\right).
\end{equation}

With these modifications our EOM become
\begin{equation}
\label{N4}
D_t u_i(\vec x,t)\,=\,g_i(\vec x,t)
\end{equation}
and
\begin{equation}
\label{N5}
D^2_t \left(\frac{g_i}{n}\right)(\vec x,t)\,=\,\frac{1}{2n(\vec x,t)}\,\partial_i(g_k(\vec x,t))^2,
\end{equation}
where
\begin{equation}\label{N6}
D_t\,=\,\partial_t\,+\,u_k(\vec x,t)\partial_k
\end{equation}
and
\begin{equation}
\label{N7}
\vec u(\vec x,t)\,=\,\dot{\vec{x}}(\vec \xi,t)\vert_{\vec \xi=\vec \xi(\vec x,t)}.
\end{equation}

To complete the set of our equations we have to include also the continuity equation for the 
particle density 
\begin{equation}
\label{N8}
\partial_t  n\,+\,\partial_k(\,n\,u_k)\,=\,0,
\end{equation}
which we generate by introducing a Lagrange multiplier field $\theta$. Then the complete
Lagrangian for the dark sector, in the Eulerian formulation, takes the form
\begin{equation}
\label{N9}
L\,=\,-\frac{1}{4\pi G}\int \,d^3x\left(g_i(D_tu_i-\frac{1}{2}g_i)\,+\,\theta(\partial_t n+\partial_i(u_in)\right).\end{equation}

Note that varying the Lagrangian with respect to the darkon density $n$ gives us the equation for
the lagrangian multiplier field $\theta$:
\begin{equation}
\label{N10}
D_t\,\theta\,=\,0.
\end{equation}
Furthermore, (\ref{N5}) is obtained from the Lagrangian (\ref{N9}) in two steps. By varying $u_i$ we obtain the EOM
\begin{equation}
\label{N10aa}
D_t g_i\,=\,g_k\partial_i u_k\,-\,g_i\partial_k u_k\,-\,n\partial_i \theta.
\end{equation}
Then we apply $D_t$ to (\ref{N10aa}) and obtain, by using the other EOM,  eq.(\ref{N5}).
Note also that in our solution (\ref{N2}) of the Poisson equation for the gravitational field 
\begin{equation}
\label{N10a}
\partial_i g_i\,=\,-4\pi G \partial_i(nq_i)
\end{equation}
 we have neglected a possible contribution to $g_i$ which is a curl
of another vector field $\vec A(\vec x,t)$.  The neglect of such term ({\it ie} putting $\vec A=0$)
is not a serious defect of our modified model as the choice (\ref{N2}) leads to the most general radial symmetric solution of the Poisson equation (\ref{N10a}) and so to the same cosmology as discussed in \cite{PRD}. 

Let us summarise our results: We have successfully reached our main aim of reducing the phase space to the physical one while keeping the cosmological solutions derived in \cite{PRD}. However, we have had to pay a prize for this as now
\begin{itemize}
\item the gravitational force is not automatically given by a gradient of a potential,
\item the EOM (\ref{N10aa}) breaks the symmetry with respect to arbitrary time-dependent translations:
\begin{equation}
\label{SS1}
x_i\,\rightarrow x_i'\,+\,a_i(t),
\end{equation}
as, under this symmetry, the fields $A$ should transform as $A\,\rightarrow\,A'$ with
$$u_i'(\vec x\,',t)\,=\,u_i(\vec x,t)\,+\,\dot a_i(t),$$
\begin{equation}
\label{SS2}
g_i'(\vec x\,',t)\,=\,g_i(\vec x,t)\,+\,\ddot a_i(t),
\end{equation}
$$n'(\vec x\,',t)\,=\,n(\vec x,t)\qquad \hbox{and}\qquad \theta'(\vec x\,',t)\,=\,\theta(\vec x,t).$$
Transformations with $a_i(t)$ being at most linearly dependent on $t$ belong to the group of Galilei transformations, whereas higher power ones (with $\ddot a\ne 0$) are gauge transformations which represent Einstein's equivalence principle.
\end{itemize}

It turns out that both problems can be solved by a further reduction of the dimension of the phase space. To do this let us assume that the velocity field $\vec u$ is curl-free, {\it i.e.}
\begin{equation}
\label{SS3}
\vec u\,=\,\vec \nabla u.
\end{equation}
Note that the restriction (\ref{SS3}) does not change the cosmological solutions. 
Next we
insert (\ref{SS3}) into the Lagrangian (\ref{N9}) and obtain
\begin{equation}
\label{SS4}
L\,=\,-\frac{1}{4\pi G}\int d^3x\left(g_i\partial_i(\partial_tu+\frac{1}{2}(\partial_ku)^2)\,-\,\frac{1}{2}g_i^2\,+\,\theta(\partial_tn+\partial_i(n\partial_iu))\right).
\end{equation}

Varying $g_i$ we obtain again the EOM (\ref{N4}) but with (\ref{SS3}) the left hand side of (\ref{N4})
becomes the gradient of a scalar and so $\vec g$ is derivable from a scalar potential $\phi$:
\begin{equation}
\label{SS5}
\vec g\,=\,-\vec \nabla \phi,
\end{equation}
where $\phi$ is given by the Bernoulli equation
\begin{equation}
\label{SS6}
-\phi\,=\,\partial_t u\,+\,\frac{1}{2}(\partial_k u)^2.
\end{equation}

Varying the velocity potential $u$ we obtain, instead of (\ref{N10aa}) the EOM
\begin{equation}
\label{SS7}
\partial_t(\partial_ig_i)\,+\,\partial_i(\partial_kg_k\partial_iu)\,+\,\partial_i(n\partial_i\theta)\,=\,0,
\end{equation}
which is nothing else than the divergence of (\ref{N10aa}) if we take (\ref{SS3}) and (\ref{SS5})
into consideration. The EOM (\ref{N8}), (\ref{N10}), (\ref{SS6}) and (\ref{SS7}) now constitute a system 
of four equations for the fields $n$, $\theta$, $\phi$ and $u$. 

It is easy to see that (\ref{SS7}) is now invariant under the set of transformations (\ref{SS1},\ref{SS2}). Note that the other equations (\ref{N4}), (\ref{N8}) and ({\ref{N10}) possess this
invariance from the very beginning. So we have saved Einstein's equivalence principle for our modified model.

\subsection{Some Comments}

\begin{itemize}
\item The Hamiltonian $H$ is according to (\ref{SS4})  given by
\begin{equation}
\label{N14}
H\,=\,\frac{1}{4\pi G}\,\int\,d^3x\left(g_i\frac{1}{2}(\partial_i(\partial_ku)^2-g_i)\,+\,\theta \partial_i(\partial_iu\,n)\right). \end{equation}
On the subspace of static solutions of the EOM we have from (26), resp. (34)
\begin{equation}
\label{NN1}
\partial_i \theta\,=\,0,\qquad \hbox{resp.}\quad \frac{1}{2}\partial_i(\partial_ku)^2\,=\,g_i
\end{equation}
and so we obtain 
\begin{equation}
\label{NN2}
H\,=\,\frac{1}{8\pi G}\,\int d^3x\,g_i^2\,>\,0,
\end{equation}
{\it i.e.} the energy of the darkon fluid is positive definitive.
But for the time-dependent solutions we can always find initial conditions
that correspond to $H$ being negative. However, as we argued in (\cite{PRD}, section IV) this does not lead to any instability.

\item The EOM (\ref{SS7}) is a continuity equation with a source term for \break \hfil $\partial_ig_i\,=\,-\triangle \phi$. 

Its important property, when compared with the standard nonrelativistic Poisson equation 
for $\phi$, is its dynamical nature {\it ie} the time development of $\triangle \phi$ is determined
by its initial data.

\item  Note that the EOM do not contain Newton's gravitational constant. The Lagrangian contains it only
as a common factor. Moreover, our modified Lagrangian (\ref{SS4}) does not split into two parts
involving a free and an interacting term. Only the EOM (\ref{N4}) (resp. (\ref{SS6})) show such a splitting. Clearly this strange property is a consequence of phase-space reduction due to (\ref{N3}). We know only one other physical system possessing this property: the interacting Chaplygin 
gas if one eliminates the particle density from the Lagrangian (see \cite{Jackiw}, section 2.1, item(i)).
 
\end{itemize}

\section{Zero mass Galilean symmetry}

The ``darkons" introduced in \cite{PRD} are free massless Galilean particles. 
As such they are a dynamical realisation of the unextended $D=(3+1)$ Galilei algebra 
(for details see \cite{us}). This time the situation 
is different. The easiest way to observe this is to go to the Lagrange formulation. Then using 
the equations derived before we note that the Lagrangian is given by
\begin{equation}\label{N13}
L\,=\,-\frac{n_0}{4\pi G}\int d^3\xi\,\left(\frac{g_i}{n}\right)(\vec x(\vec \xi,t),t)\left(\ddot x_i(\vec \xi,t)-g_i(\vec x(\vec \xi,t),t)\right) \,-\end{equation}
$$-\,\frac{1}{8\pi G}\int d^3x\,g_i^2(\vec x,t).$$

This shows that $L$ does not possess a particle limit. Darkons exist only as a self-gravitating
fluid. The questions therefore arises which symmetry algebra is dynamically realised by such a fluid.

In this section we are going to consider the Galilei algebra leaving the possible conformal
extension to the next section. So, let us start with the Hamiltonian $H$ (generator of time
translations) which has already been  given by (\ref{N14}).

We can now construct the conserved generators for space translations and Galilean boosts.
According to Noether's theorem we obtain for the generator $P_i$ of space translations
\begin{equation}
\label{N15}
P_i\,=\,\int d^3x\,np_i,
\end{equation}
where 
\begin{equation}
\label{N16}
(np_i)(\vec x,t)\,=\,-\frac{1}{4\pi G}(\partial_kg_k\partial_iu\,+\,n\partial_i\theta)
\end{equation}
and the Galilean boosts are generated by
\begin{equation}
\label{N17}
K_i\,=\,tP_i\,+\,\frac{1}{4\pi G}\int \,d^3x x_i\partial_kg_k.
\end{equation}
 By using the EOMs it is easily seen that both 
$P_i$ and $K_i$ are conserved.
However, the Poisson bracket of these generators vanishes
\begin{equation}
\label{N18}
\{K_i,\,P_j\}\,=\,0
\end{equation}
as is easily inferred from the property of $P_i$ as the generator of space translations
\begin{equation}
\label{N18a}
\{P_i,\,A(\vec x,t)\}\,=\,\partial_i A(\vec x,t)
\end{equation}
for any fluid field $A$. 

The vanishing of the right hand side of (\ref{N18}) tells us that the total mass of our darkon fluid is zero.

If we consider, in addition, the conserved angular momentum (generator of space rotations)
\begin{equation}
\label{N18aa}
\vec J\,=\,\int d^3x\vec x\times n\vec p
\end{equation}
we clearly see that the ten generators $H,P_i,K_i$ and $J_i$ satisfy the unextended, {\it i.e.} zero-mass  Galilei 
algebra.

\section{Anisotropic scaling}
Let us note that we have as a conserved dilation generator
\begin{equation}
\label{N19}
D\,=\,Ht\,-\,\frac{1}{5}\frac{1}{4\pi G}\,\int d^3x\,u\partial_i g_i\,-\,\frac{3}{5}\int d^3x\,x_inp_i,
\end{equation}
which leads to the Poisson bracket with the translation generator satisfying  $\{P_i,D\}=\frac{3}{5}P_i$
and thus corresponds to the dynamical exponent $z=\frac{5}{3}$ {\it ie} time and space coordinates
scale as (see \cite{us})
\begin{equation}
\label{N20}
x^{\star}_i\,=\,\lambda x_i,\qquad t^{\star}\,=\,\lambda^{\frac{5}{3}}t.
\end{equation}
To work out the scale dimension $z_A$ of a field $A(\vec x,t)\in (n,u,\phi,\theta)$ defined by
\begin{equation}
\label{N21}
A^{\star}(\vec x,t)\,=\,\lambda^{z_A}A(\vec x^{\star},t^{\star}),
\end{equation}
we look at the scale invariance of the action.

From the requirement that all terms in the Lagrangian scale in the same way we find
\begin{equation}
\label{N22}
z_u\,=\,z-2,\quad z_{\phi}\,=\,2z-2,\quad \hbox{and}\quad z_{\theta}+z_n\,=\,3z-2.\end{equation}
However, the measure of the integration in the action scales as $\lambda^{-3-z}$ and so again
$z\,=\,\frac{5}{3}$ and therefore we find that 
\begin{equation}
\label{N23}
z_u\,=\,-\frac{1}{3},\quad z_{\phi}\,=\,\frac{4}{3},\quad z_{\theta}+z_n\,=\,3.
\end{equation}
Note that (\ref{N23}) leaves $z_n$ undetermined because $D$ is determined only modulo a term $\sim Q$, where $Q$ is the conserved charge $Q=\int d^3x\,n\theta$. In (\ref{N19}) we have fixed this term in such a way that we have 
 scale invariance of the particle number 
\begin{equation}
\label{N24}
N\,=\,\int d^3x\,n(\vec x,t)
\end{equation}
and so we get $z_n=3$ and in consequence $z_{\theta}=0$.
We conclude that the symmetry algebra of our darkon fluid becomes the expansion-less conformal 
Galilean-type algebra with $z=\frac{5}{3}$ (cf. \cite{us}).


To derive the form of our EOM, either for the scale invariant solutions or for cosmological perturbations, it is convenient to perform the transformation

\begin{equation}
\label{M25}
(\vec x,t)\qquad \rightarrow \qquad (\vec w,t)
\end{equation}
with $\vec w=\vec x\left( \frac{t}{t_0}\right)^{-\frac{3}{5}}$ and let
\begin{equation}
\label{M26}
A(\vec x,t)\,=\,\left(\frac{t}{t_0}\right)^{-\frac{3z_A}{5}}\tilde A(\vec w,t).
\end{equation}

As $\vec w$ is scale invariant we see from (\ref{N21}) that $\tilde A$ scales as
\begin{equation}
\label{M27}
\tilde A^{\star}(\vec w,t)\,=\,\tilde A(\vec w,\lambda^{\frac{5}{3}}t)
\end{equation}
and so $A$ is scale invariant ($A=A^{\star}$) if
\begin{equation}
\label{M28}
\partial_t \tilde A(\vec w,t)\,=\,0.
\end{equation}
But, due to (\ref{M26}), the partial derivatives of $A$ transform as
\begin{equation}
\label{M29}
\partial_i A(\vec x,t)\,=\,\left(\frac{t}{t_0}\right)^{-\frac{3}{5}(1+z_A)}\,\partial_i \tilde A(\vec w,t)
\end{equation}
and
\begin{equation}
\label{M30}
\partial_t A(\vec x,t)\,=\,\left(\frac{t}{t_0}\right)^{-\frac{3z_A}{5}}\,\left(\partial_t -\frac{3}{5t}(w_k\partial_k+z_A)\right)\tilde A(\vec w,t).
\end{equation}
Hence, in terms of $\tilde A$ the EOM (\ref{SS6}), (\ref{N8}), (\ref{N10}) and (\ref{SS7}) take the form
\begin{equation}
\label{M31}
(t\partial_t\,-\,\frac{3}{5}w_k\partial_k\,+\,\frac{1}{5})\tilde u\,+\,\frac{t_0}{2}(\partial_k\tilde u)^2\,=\,-t_0\tilde \phi
\end{equation}
\begin{equation}
\label{M32}
t\partial_t\tilde n\,+\, \partial_k\left((t_0\partial_k\tilde u-\frac{3}{5}w_k)\tilde n\right)\,=\,0,
\end{equation}
\begin{equation}
\label{M33}
t\partial_t \tilde \theta\,+\,(t_0\partial_k\tilde u-\frac{3}{5}w_k)\,\partial_k\tilde \theta\,=\,0
\end{equation}
and
\begin{equation}
\label{M34}
(t\partial_t-\frac{3}{5}w_k\partial_k-2)\triangle \tilde \phi\,+\,t_0\partial_i(\triangle \tilde \phi\partial_i\tilde u)\,=\,t_0\partial_i(\tilde n\partial_i\tilde \theta).
\end{equation}

It is striking that these EOM contain an explicit time dependence in the form of the term $t\partial_t$. To give them the standard 1-st order form it is convenient to perform the transformation
\begin{equation}
\label{M34a}
t\,\rightarrow\, \tau\,=\,\log\left(\frac{t}{t_0}\right), 
\end{equation}
or, for the derivatives, $t\partial_t=\partial_{\tau}$.

Then the EOM (\ref{M31}-\ref{M34}) are derivable from the Lagrangian (note that $\tilde g_i$ and not $\tilde \phi$ is the relevant variable in $\tilde L$)
\begin{equation}
\label{M34b}
\tilde L\,=\,-\frac{1}{4\pi G}\int d^3w\left(\tilde g_i(\partial_{\tau}+(t_0\partial_k\tilde u-\frac{3}{5}w_k)\partial_k-\frac{2}{5})\partial_i\tilde u\right. \,-
\end{equation}
$$\left.-\frac{t_0}{2}\tilde g^2_i\,+\,\tilde \theta(\partial_{\tau}\tilde n\,+\,\partial_k(
(t_0\partial_k\tilde u-\frac{3}{5}w_k)\tilde n)\right)$$
which may be obtained from the original action by inserting the transformation (\ref{M25},\ref{M26}) and (\ref{M34a}).
The corresponding Hamiltonian $\tilde H$ can be read off from (\ref{M34b}) by writing it as
\begin{equation}
\label{M34c}
\tilde L\,=\,-\frac{1}{4\pi G}\int d^3w(\tilde g_i\partial_{\tau}\partial_i\tilde u\,+\,\tilde \theta\partial_{\tau}\tilde n)\,-\,\tilde H.
\end{equation}
Now we have to be cautious: The Hamiltonian $\tilde H$ is {\bf not} the transformed form of $H$ (eq. (\ref{N14})) but it comes from inserting the above mentioned transformations into the dilation generator $D$ (\ref{N19}). The reason for this replacement of $H$ by $D$ is an immediate consequence of the transformation (\ref{M34a}) which leads in (\ref{M27}) to
\begin{equation}
\label{M34d}
\tilde A^{\star}(\vec w,\tau)\,=\,\tilde A(\vec w,\tau+\frac{5}{3}\log \lambda)
\end{equation}
{\it i.e.} time rescalings become time translations (both to be taken at fixed $\vec w$).

\skip 0.5cm

Note that for scale invariant solutions, characterised by (\ref{M28}) eq. (\ref{M33})
turns into
\begin{equation}
\label{M35}
\left(t_0\partial_k\tilde u\,-\,\frac{3}{5}w_k\right)\partial_k\tilde \theta\,=\,0.
\end{equation}
Thus we have to distinguish three different possibilities
\begin{itemize}
\item {1.}
\begin{equation}
\label{M36}
\tilde u_0(\vec w)\,=\,\frac{3}{10}t_0^{-1}w^2,
\end{equation}
where $w=\vert \vec w\vert$.
Then (\ref{M32}) is identically fulfilled and from (\ref{M31}), resp. (\ref{M34}) we obtain
\begin{equation}
\label{M37}
\tilde \phi_0\,=\,\frac{3}{25}t_0^{-2}w^2,
\end{equation}
and, resp., 
\begin{equation}
\label{M38}
\partial_k(\tilde n_0(\vec w)\partial_k \tilde \theta_0(\vec w))\,=\,-\frac{18}{125}t_0^{-3}.
\end{equation}

In the next section we shall show that the choice $\tilde n_0(\vec w)=$const. leads to the cosmological solutions for the early universe.
\item{2.}
\begin{equation}
\label{M39}
\partial_k\tilde \theta\,=\,0
\end{equation}
but $t_0{\vec\nabla \tilde u}-\frac{3}{5}\vec w\ne0$.
Then (\ref{M32}) has no radially symmetric solutions valid for all
$w\ge0$. To prove this statement we note that the equation $\partial_k(w_kf(w))=0$ has, for nonvanishing $w$, the unique solution $f(w)\sim w^{-3}$ leading to $\partial_k(w_kf(w))\sim \delta(\vec w)$. 
We have not so far succeeded in finding, analytically, the anisotropic solutions.
\item {3.}
\begin{equation}
\label{M40}
\left(t_0 \vec{\nabla}\tilde u\,-\,\frac{3}{5}\vec w\right) \perp \vec \nabla \tilde \theta.
\end{equation}
The discussion of (\ref{M40}) lies outside the scope of the present paper.

\end{itemize}

\section{Cosmological solutions and their perturbations at early times}


The cosmologically relevant solutions, as derived and discussed in \cite{PRD}, are given by
\begin{equation}
\label{N31}
n_0(\vec x,t)\,=\,\frac{3D}{4\pi a^3(t)},
\end{equation}
\begin{equation}
\label{N32}
\vec u_0(\vec x,t)\,=\,\frac{\dot a}{a}\vec x,
\end{equation}
and
\begin{equation}
\label{N33}
\vec g_0(\vec x,t)\,=\,-\frac{3DG}{a^3}\,g(a)\,\vec x,
\end{equation}
where $D$ is a positive constant.

The unknown cosmic scale factor $a(t)$ and the function $g(a)$ obey the following set
of differential equations:
\begin{equation}
\label{M63}
\ddot a\,=\,-\frac{3DG}{a^2}g(a)
\end{equation}
and 
\begin{equation}
\label{N34}
\dot g(a(t))\,=\,\frac{\beta}{a^2},
\end{equation}
where $\beta$ is a constant. Then we have


\begin{equation}
\label{N35}
\theta_0(\vec x,t)\,=\,2\pi G \beta \frac{r^2}{a^2}.
\end{equation}

The last expression (\ref{N35}) comes from the observation that for the cosmological solutions
the equation for $\theta$ reduces to
\begin{equation}
\label{N36}
\left( \partial_t\,+\,\frac{\dot a}{a}\,x_k\partial_k\right) \theta_0\,=\,0,
\end{equation}
which has (\ref{N35}) as its solution. The overall constant comes from inserting the expressions above
into the equation for $g_i$ {\it ie} (\ref{SS7}).

Comparing (\ref{N31}-\ref{N35}) with the scale invariant solutions of case 1. (\ref{M36}-\ref{M38}) we note that we have agreement if
\begin{equation}
\label{M41}
\tilde n_0(\vec w)\,=\,\frac{3D}{4\pi},
\end{equation}
and
\begin{equation}
\label{M42}
a(t)\,=\,\left(\frac{t}{t_0}\right)^{\frac{3}{5}},
\end{equation}
where $t_0$ becomes the cosmological time at present and
\begin{equation}
\label{M43}
\beta\,=\,-\frac{2}{125}(DG)^{-1}t_0^{-3}.
\end{equation}
According to \cite{PRD} eq. (\ref{M42}) gives the cosmic scale factor $a(t)$ at early times, {\it i.e.} 
our cosmological solutions are scale invariant for times before structure formation sets in.
Therefore the scaling law (\ref{N20}) is not applicable to the relations between microscopic scales 
and galactic ones (see \cite{newten}).

\subsection{Perturbations}
Given the cosmological solutions we can now consider their perturbations.

To do this we take any fluid field $\tilde A\in (\tilde n,\tilde u_i,\tilde g_i,\tilde \theta)$ and put
\begin{equation}
\label{N39}
\tilde A(\vec w,t)\,=\,\tilde A_0(\vec w)\,+\,\delta \tilde A(\vec w,t)
\end{equation}
where the first term in (\ref{N39}) denotes the cosmological background solution and the second its 
perturbation. Note that $\delta \tilde n$ is proportional to the fractional density 
perturbation $\delta \tilde n\sim \frac{\delta n}{n_0}$ and $\vec w$ is the coordinate comoving
with the background.

Putting the expressions (\ref{N39}) into the Lagrangian (\ref{M34b}) and keeping terms quadratic in 
perturbations gives us (we abreviate $\hat A=\delta\tilde A$)
\begin{equation}
\label{N39a}
\hat L\,=\,-\frac{1}{4\pi G}\int d^3w\left( \hat g_i(\partial_{\tau}+\frac{1}{5})\partial_i\hat u\,-\,
\frac{t_0}{2}\hat g^2_i\,+\,\frac{9}{25}t_0^{-1}(\partial_i\tilde u)^2\,+\right.
\end{equation}
$$+\left.\hat\theta(\partial_{\tau}\hat n\,+\,\frac{3D}{4\pi}t_0\triangle\hat u)\,+\,\frac{8\pi}{125Dt_0^2}
\hat nw_k\partial_k\hat u\right).$$

It turns out that this Lagrangian is invariant with respect to the rescaling of the comoving space coordinates $\vec w$ when the fields $\hat A\in (\hat n, \hat \theta, \hat g_i, \hat u)$ transform as
\begin{equation}
\label{N39b}
\hat A(\vec w,\tau)\,\rightarrow\, \hat A^{\star}(\vec w,\tau)\,=\,\lambda^{\eta_A}\hat A(\lambda \vec w,\tau)
\end{equation}
with 
\begin{equation}
\label{N39c}
\eta_u\,=\,\frac{1}{2},\quad \eta_g\,=\,\frac{3}{2},\quad \eta_n\,=\,\frac{5}{2}\quad\hbox{and}\quad\eta_{\theta}\,=\,\frac{1}{2}.
\end{equation}

By Noether's theorem we obtain for the generator $D_s$ of this symmetry
\begin{equation}
\label{N39d}
D_s\,=\,\int d^3w\,\left(-\partial_i\hat g_i(\frac{1}{2}\,+\,w_k\partial_k)\hat u\,+\,\hat \theta(\frac{5}{2}\,+\,w_k\partial_k)\hat n\right).
\end{equation}

The EOM following from the Lagrangian (\ref{N39a}) are

\begin{equation}
\label{N48}
(\partial_{\tau}\,+\,\frac{1}{5})\hat u\,=\,-t_0\hat \phi,
\end{equation}
\begin{equation}
\label{N47}
\partial_{\tau} \hat n\,+\,\frac{3D}{4\pi}t_0\triangle\hat u\,=\,0,
\end{equation}
\begin{equation}
\label{N50}
\partial_{\tau}\hat \theta\,=\,\frac{8\pi}{125 D t_0^2}w_k\partial_k\hat u,
\end{equation}
and
\begin{equation}
\label{N49}
(\partial_{\tau}-\frac{1}{5})\triangle \hat \phi\,+\,\frac{18}{25t_0}\triangle \hat u\,-\,\frac{3D}{4\pi}t_0\triangle \hat \theta\,+\,\frac{8\pi}{125 D t_0^2}\partial_i(w_i\hat n)\,=\,0.
\end{equation}

Applying $\partial_{\tau}$ to (\ref{N49}) it  is possible to eliminate $\hat n$ and $\hat \theta$. One then obtains
\begin{equation}
\label{N51}
\partial_{\tau}(\partial_{\tau}-\frac{1}{5})\triangle \hat \phi\,-\,\frac{18}{25}\triangle \hat \phi\,-\,\frac{12}{125t_0}(4+w_i\partial_i)\triangle \hat u\,=\,0.
\end{equation}


With the ansatz
\begin{equation}
\label{N52}
\hat u(\vec w,\tau)\,=\,e^{\alpha \tau}u_{1}(\vec w),
\end{equation}
for some $\alpha$ we find  from (\ref{N51}) with (\ref{N48}) that
\begin{equation}
\label{N54} 
(a_1\,+\,a_2w\frac{\partial}{\partial w})\triangle u_1(\vec w)\,=\,0,
\end{equation}
where the constants $a_{1,2}$ are given by
\begin{equation}
\label{N55}
a_1\,=\,\alpha(\alpha^2-\frac{19}{25})\,+\,\frac{30}{125},\qquad a_2\,=\,\frac{12}{125}
\end{equation}


Let us now consider the particular case when $ u_1$ is invariant with respect to the space rescalings (\ref{N39b}). Then we have in spherical coordinates
\begin{equation}
\label{P60f}
 u_1(\vec w)\,=\,w^{-\frac{1}{2}}f(\theta,\varphi),
\end{equation}
which gives us from (\ref{N54}) an equation for $\alpha$
\begin{equation}
\label{P60g}
a_1\,-\,\frac{5}{2}a_2\,=\,0
\end{equation}
and therefore
\begin{equation}
\label{PPa}
\alpha(\alpha^2-\frac{19}{25})\,=\,0
\end{equation}
with the solutions
\begin{equation}
\label{P60h}
\alpha\,=\,0, \qquad \alpha=\pm \sqrt{\frac{19}{25}}
\end{equation}

Expanding $f(\theta,\varphi)$ into spherical harmonics $Y_{l,m}$  we obtain for the growing mode of $\hat n$
\begin{equation}
\label{P60i}
\hat n(\vec w,t)\,\sim\, \left(\frac{t}{t_0}\right)^{0.8718}w^{-\frac{5}{2}}Y_{l,m}.
\end{equation}



As the EOM (\ref{N48}-\ref{N49}) are linear the general solution for $\hat n$ is a linear combination of the partial solutions  (\ref{P60i}) with their relative weight determined by the initial conditions.

Clearly, the existence of structures, such as galaxies and stars, in the universe suggests that the perturbations, at early times, were not radially symmetric (cf \cite{weinberg}).

\section{Adding higher-order derivative terms}

A generalization of the EOM (\ref{N4}) which preserves gauge invariance is obtained by adding terms containing spatial derivatives of $g_i$. Such terms are generated by adding to the Lagrangian (\ref{SS4}) terms of the form (we consider only terms which are quadratic in $g_i$)
\begin{equation}
\label{N61}
L'\,=\,-\frac{1}{8\pi G}\int d^3x\,(K_1(\partial_ig_i)^2\,+\,K_2(\partial_k\partial_ig_i)^2\,+\,...).
\end{equation}
The first trem in $L'$ corresponds, for $g_i=-\partial_i \phi$, to the space-part of the interaction term 
for a Lifshitz scalar field (see \cite{Horava} and the literature cited therein).  Clearly $[K_n]=(\hbox{length})^{2n}$.

Note that the EOM remain the same as before except for the Euler equation which now becomes
\begin{equation}
\label{N62}
\partial_i(\partial_tu\,+\,\frac{1}{2}(\partial_ku)^2)\,=\,g_i\,-\,K_1\partial_i(\partial_kg_k)
\,+\,K_2\partial_i\triangle\partial_kg_k\,+\,....
\end{equation}
Note that
\begin{itemize}
\item The terms proportional to the $K_n$ do not contribute to the cosmological solution 
given by $g_k(\vec x,t)=x_kg(t)$.
\item Whether the $K_n$ vanish or not the system of equations does not possess static radially
symmetric solutions.
\item Note that (\ref{N61}) scales differently than the $g_i^2$ term in (\ref{SS4}).
      To be specific suppose that we consider only the first term in (\ref{N61}) as the interaction term. Then 
we find that our fields have the scale dimensions 
$z_u=\frac{1}{3}$, 
$z_g=\frac{5}{3}$ and the dynamical exponent is given by $z=\frac{7}{3}$.  This suggests that when we have both
interaction terms we have a model which has a dynamical exponent $z=\frac{7}{3}$ at short distances and
$z=\frac{5}{3}$ at long ones.
\end{itemize}

\section{Conclusion and Outlook}

The main aims of this paper involved:
\begin{itemize}
\item the elimination of the unphysical degrees of freedom in the original formulation of the model
\cite{PRD} for the description of a self-gravitating darkon fluid,
\item the demonstration that this darkon fluid exhibits the zero mass Galilean symmetry and, in consequence, possesses anisotropic scaling symmetry,
\item the exploitation of the anisotropic scaling symmetry for the formulation of cosmological 
perturbations within the dark sector of the universe.
\end{itemize}


In order to obtain results that could be compared with cosmological data we have to add baryonic matter and the higher spatial-derivative terms 
as  further perturbations. Then, however, everything becomes more complicated as any of these contributions break the dilatational symmetry of our EOM. Thus we will leave such an extension of our present work 
to a subsequent paper.

Our model possesses the symmetry with respect to time-dependent space translations
\begin{equation}
\label{Z1} 
x_i\,\rightarrow \,x_i'\,=\,x_i\,+\,a_i(t).
\end{equation}

We plan to try to generalize our model so that it comes as close as possible to General Relativity which possesses symmetry with respect to arbitrary space-time diffeomorphisms ($Diff$). In a 
nonrelativistic setting we have therefore to consider the subgroup of $Diff$ which leaves the Newtonian time fixed (modulo reparametrizations). 
\begin{equation}
\label{Z2}
x_i\,\rightarrow\,x_i'\,=\,x_i\,+\,a_i(\vec x,t),\qquad t\,\rightarrow t'\,=\,f(t).
\end{equation}
In order to do this we have to start by `covariantizing' the time derivative terms in our Lagrangian (\ref{SS4}). Such a `covariantizing' procedure has been worked out already some time ago for massive point particles in (1+1) \cite{nine}, respectively (2+1) \cite{ten} dimensions. This project is currently being studied and we hope to present our results soon.

\end{document}